# Magnetoelectric coupling in Haldane spin chain system, $Dy_2BaNiO_5$

Kiran Singh[1], Tathamay Basu[1], S. Chowki[2], N. Mahapotra[2], K. Iyer[1], P. L. Paulose[1] and E.V. Sampathkumaran[1]

[1]Tata Institute of Fundamental Research, Homi Bhabha Road, Colaba, Mumbai- 400005, India

[2]School of Basic Sciences, Indian Institute of Technology-Bhubaneswar, Bhubaneswar, 751013, India

## Abstract:

We report the results of various measurements, namely magnetization, complex dielectric permittivity and electric polarization ($P$) on $Dy_2BaNiO_5$ as a function of temperature ($T$) and magnetic-field ($H$), apart from heat-capacity ($C$), with the primary motivation of exploring the existence of magnetoelectric ($ME$) coupling among Haldane spin-chain systems. The $M(T)$ and $C(T)$ data establish long range magnetic ordering at 58K. The most noteworthy observations are: (i) Distinct anomalies are observed in dielectric constant ($\varepsilon'$) *vs* $T$ and loss ($tan\delta$) *vs* $T$ at different temperatures, i.e. 12.5, 30, 50 and 58K; at low temperatures, three magnetic-field-induced transitions are observed in $\varepsilon'$ *vs* $H$ at 6, 40 and 60 kOe. These transition temperatures and critical magnetic fields track those obtained from magnetization data. This establishes the existence of strong magnetoelectric coupling in this compound. (ii) Correspondingly, electric polarization could be observed as a function of $T$ and $H$ in the magnetically ordered state, thereby indicating magnetism-induced ferroelectricity in this compound; this result suggests that this compound is a possible new multiferroic material among spin = 1 (nickel containing) compounds—with successive magnetic transitions and strong magnetoelectric coupling.

**PACS number(s):** 75.85.+t, 75.30.Kz, 77.80.B-, 75.40.Cx

# I.     INTRODUCTION

The magnetoelectric (*ME*) materials where the magnetic (electric) properties can be tuned by electric (magnetic) field have gained a great attention because of their importance for fundamental science and technology. After the discovery of multiferroicity and *ME* coupling in TbMnO$_3$ [1], many new multiferroics with strong *ME* coupling have been discovered with different microscopic origins [2, 3]. In the vast growing field of multiferroicity, many of the magnetically driven ferroelectrics are geometrically frustrated magnetic materials.

In this article, we focus our attention on the spin-chain systems of the type $R_2$BaNiO$_5$ (*R*= Rare-earths) [3-19], which do not exhibit geometrically frustrated magnetism. Among spin-chain systems, rare earth nickelates (spin, $S = 1$) have drawn great interest in recent years due to the observation of Haldane gap [20 ] even in the long range magnetically ordered state [4, 11, 17-19] for magnetic moment containing *R*-members. The Y$_2$BaNiO$_5$ ($S = 1$) is the first one in this series originally shown to exhibit Haldane gap $\Delta_H = 10$ meV, however without long range ordering down to 100 mK [5-7, 9, 10]. The long range magnetic ordering is missing in Y$_2$BaNiO$_5$ because the intrachain interaction is stronger than interchain interaction. The Ni chain is isolated by Ba$^{2+}$ and Y$^{3+}$ (which are non-magnetic) and hence no long range magnetic ordering has been observed.  Therefore, the observation of long range antiferromagnetic (*AFM*) ordering for other members implies significant role of *R*-Ni magnetic interactions [4, 12-15].

All $R_2$BaNiO$_5$ compounds crystallize in orthorhombic structure (*Immm*) [15]. However, heavier rare earths (Tm, Yb and Lu) are dimorphic and depending on the synthetic conditions, these can also form in the well known '*green phase*' having *Pnma* space group [13]. As compared to regular octahedra, the NiO$_6$ octahedra in this family is strongly distorted. The Ni-O apical distance is 1.88Å, whereas the corresponding distance in the basal plane is 2.18Å.

Moreover, the O-Ni-O bond angle is 78°, much smaller than that for the regular octahedra [15]. Since the interchain to intrachain interaction ratio is very small (~ $10^{-2}$), one-dimensional (1*D*) behavior apparently is perceived to persist even in the magnetically ordered state [17, 18]. It is well-known that, for small values of *S*=1/2 or 1, strong quantum fluctuations occur at low temperatures (*T*), leading to exotic ground states [21]. Therefore, $R_2$BaNiO$_5$ series is considered to be a rare one showing both classical and quantum spin dynamics in the magnetically ordered state [18].

The magnetism of $R_2$BaNiO$_5$ compounds has been studied in detail by neutron diffraction [15]. For magnetic-moment containing *R*, it is believed that there are two magnetic anomalies, one as revealed by a maximum in magnetic susceptibility ($\chi$) *vs T* at a temperature, $T_{max}$, and another at a higher temperature ($T_N$) arising from 3-dimentional (3*D*) long range magnetic ordering. For the whole series, $T_N$ is greater than $T_{max}$. In the earlier stages of investigation of this series, $T_{max}$ was considered to be due to 3*D* antiferromagnetic (*AFM*) ordering [22]; subsequent neutron diffraction studies evidenced that both $R^{3+}$ and Ni$^{2+}$ magnetic moments actually order well above the $T_{max}$ [4]. The magnetic structure is commensurate *k(1/2, 0, 1/2)* in the absence of an external magnetic-field (*H*). It may be noted that most of the compounds of this family have been known to exhibit only two magnetic-field-induced transitions below $T_{max}$ [12, 16].

Turning to dielectric properties, recently, Chen et al. [23] have studied the dielectric behavior of Y$_2$BaNiO$_5$. Above 80K, this compound is not a good insulator [6] and very high dielectric permittivity ($\varepsilon' = 10^4$ at room temperature) was observed. In the case of Ho$_2$BaNiO$_5$, the linear *ME* coupling was reported [24]. However, to the best of our knowledge, there is no further report on the behavior of $\varepsilon'$ *vs T* (except for Y$_2$BaNiO$_5$) and $\varepsilon'$ *vs H* for any member of this series. As most of the $R_2$BaNiO$_5$ compounds show magnetic-field-induced transitions [12, 16] and are highly insulating at low temperatures, it would be interesting to study magnetic and dielectric properties (as a function of $T$ and $H$) to explore the existence of $ME$ coupling in these systems. In this series, $T_N$ varies with rare-earth ionic radius and maximum $T_N$ (~65K) is observed for $R$ = Tb. The value of $T_N$ for the Dy member is comparable (~58K) to that of Tb. The Dy member is unique in this series, as, at low $T$, $Ni^{2+}$ moment tends to align in positive $c$-axis, i.e., towards $Dy^{3+}$ magnetic moment [15] unlike in other members. In view of such interesting magnetic behavior, we have chosen this compound for the present investigation. Our results reveal that this compound behaves like a multiferroic material with strong $ME$ coupling. Another outcome of this work is that there are in fact three, rather than two, magnetic-field-induced transitions in this compound.

## II.    EXPERIMENTAL

Polycrystalline sample of $Dy_2BaNiO_5$ was prepared by a standard solid state reaction route as described in literature [12, 15, 25]. The formation of the compound was ascertained by room temperature X-ray powder diffraction pattern using $Cu\text{-}K_\alpha$ radiation. The Rietveld refinement also confirmed phase formation and the lattice parameters ($a$ = 3.723(1) Å, $b$ = 5.702(2) Å and $c$ = 11.208(4) Å) are in good agreement with those reported in the literature [8, 15]. The $dc$ $\chi$ measurements were carried out in the temperature interval 1.8 - 300 K in the presence of magnetic fields of 100 Oe and 5 kOe for zero-field-cooled ($zfc$) and field-cooled ($fc$) conditions using a commercial Superconducting Quantum Interference Device (SQUID, Quantum Design, USA) and isothermal magnetization ($M$) behavior was also recorded at selected temperatures with the help of a commercial Vibrating Sample Magnetometer (VSM, Quantum Design, USA).  In addition, heat-capacity ($C$) measurements were done by a relaxation method using commercial Physical Properties Measurement System (PPMS, Quantum Design). We attempted to measure temperature dependent $dc$ electrical resistivity ($\rho$) by a four-probe method in zero and 70kOe magnetic field using commercial PPMS as well as by a two-probe method (not shown here). The value at room temperature falls in the mega-Ohms-cm range and it appears to increase with decreasing temperature; below 210 K, it is found to be beyond detection limit by four-probe measurements. The magnetoresistance was found to be negliglible, as inferred from the data measured till 70 kOe. Complex dielectric permittivity was measured using Agilent E4980A LCR meter with a home-made sample-holder integrated with the PPMS. Temperature dependent complex dielectric permittivity was measured at 30 kHz at 1V ac bias during warming (1K/min). Remnant polarization ($P$) as a function of $T$ was measured with Keithley 6517A electrometer in Columbic mode. An electric field of 400kV/m was applied at 70K to align the electric dipoles and was removed after cooling to 8K. After that capacitor was discharged for 30 minutes and $P$ vs time was recorded for 5ksec to remove stray charges, if any. $P$ $vs$ $T$ was measured during warming (5K/min). For $P$ $vs$ $H$ measurements, the same procedure was followed as explained for $P$ $vs$ $T$. Isothermal magnetic-field dependence of dielectric behavior was measured at different temperatures (below and above $T_N$) at 100 kHz. The rate of change of magnetic-field in all these measurements was 100 Oe/sec.

## III.   RESULTS

### A.   Magnetic susceptibility

Temperature dependence of magnetic susceptibility is shown in Fig. 1. The $\chi(T)$ behavior obtained in 5 kOe for $Dy_2BaNiO_5$ is presented in Fig. 1a. The curves obtained for $zfc$ and $fc$ conditions overlap; therefore, we have shown only the $zfc$ curve. As the temperature is lowered, the plot of $\chi$ *vs T* shows a small kink near 58K, which is distinctly visible in the first derivative, d$\chi$/d*T* (also plotted in Fig. 1a, right y axis), arising from the onset of long range magnetic-order. This is followed by a broad hump around 38K which is attributed to the persistence of one-dimensional magnetic feature in the magnetically ordered state. In addition below $T \sim 10K$, $\chi$ increases with decreasing *T* as though there is another magnetic transition. The inverse of $\chi$ is linear above around 100 K (see Fig. 1(b)) and a Curie-Weiss fit leads to a value of -24K for the paramagnetic Curie temperature ($\theta_p$) and an effective magnetic moment ($\mu_{eff}$) of 10.73($\pm$.05) $\mu_B$/Dy atom in agreement with the reported results [15, 26, 27].

## B.    Heat-capacity

The heat-capacity of the title compound, important to infer about the onset of long range magnetic ordering, has not been reported in the literature. We present this property in the form of $C/T$ as a function of *T* in Fig. 2a. A clear peak is observed around 58K, attributable to 3*D* long range magnetic ordering. This 3*D* ordering temperature is consistent with optical absorption and neutron diffraction results [16, 27]. A convex-shaped feature around 18-38K and an upturn around 10K signal additional transitions at low temperatures. Such an additional transition at low temperature (around 10K) was also proposed by Chepurko et. al. [16] for $Er_2BaNiO_6$. In order to see these features more clearly, we have shown $d(C/T)$ *vs* $d(T)$ in Fig. 2b, which reveals an upturn below about 30K with a peak near 12K. These features are also reflected in d$\chi$/d*T* plot (Fig. 1a). We believe that such multiple features are associated with magnetism and can be related to the temperature evolution of $Dy^{3+}$ and $Ni^{2+}$ magnetic moments and Dy-Dy interactions [15]. Similar heat-capacity behavior has actually been observed for some other members of this rare-earth series [14, 28].

## C. Temperature dependence of dielectric constant and electric polarization

Figures 3a and 3b present $T$-dependent $\varepsilon'$ and loss factor ($tan\delta$) near magnetic ordering region. The anomaly at $T_N$ in $\varepsilon'$ is very weak and is difficult to see (Fig. 3a). Though one can see a change in slope at $T_N$ by plotting $d\varepsilon'/dT$ (not shown here), the existence of a transition can be clearly seen in the plot of $d^2\varepsilon'/dT^2$ ( inset 3a bottom left). Additionally, there is a weak change in the slope around 30 K; another one is also present in the range 12-15 K as shown in the inset (upper right) of Fig. 3a. The feature near 12.5 K can be seen clearly in $tan\delta$ vs $T$ (Fig. 3b, also see inset). All these features can be seen clearly in the plot of $d(tan\delta)/dT$ vs $T$ (Fig. 3b, right y-axis), including a peak at $T_N$.

In order to explore the presence of ferroelectricity, remnant electric polarization was measured as a function of $T$. The results presented in Fig. 3c show that electric polarization exists in the magnetically ordered state (below $T_N$) indicating spin-driven polarization. Open circles represent the observed data points and red line is the polynomial fit (ninth order). This polynomial fit act as a guide to eyes without any other implications. Step-like features in the observed data is due to the instrumental resolution (1pC), because the changes in $P$ are very negligible for a small interval of temperature (with the data taken after each second). The remnant $P$ is less compared to that observed for $Ho_2BaNiO_5$ [24]. In addition to the presence of polarization in the magnetically ordered state, we see distinct anomalies in other temperature ranges, viz., around 50, 38, 30 and 10K. The feature around 50K coincides with the temperature at which $\varepsilon'$ and $tan\delta$ start increasing rapidly with temperature (see Fig. 3b, right y axis) and this is related to a sudden change in the magnetism due to Dy. The feature around 38K may be associated with the rotation of $Ni^{3+}$ magnetic moments (from negative to positive direction with respect to $c$-axis); we believe that all the temperature-induced features arise from gradual rotation of Ni magnetic moments towards Dy moment (see below for further discussions on this aspect). In any case, observation of multiple features in similar temperature ranges in $d(\chi)/dT$, $C/T$, $d(C/T)/dT$, $\varepsilon'(T)$ and $P(T)$ offers direct evidence for the coupling between electric and magnetic ordering parameters.

### D. Isothermal magnetoelectric coupling effects

Magnetic-field dependent $M$, $\varepsilon'(T)$ and $P$ results are presented in Fig. 4. In this figure, panels $a$, $b$ and $c$ show isothermal magnetization behavior at 2, 5 and 12.5K. At low temperatures, say at 2 and 5K, magnetic-field-induced magnetic transitions are clearly observed. The one around 45 kOe ($H_{c2}$) and another around 60 kOe ($H_{c3}$) are consistent with earlier reports [12, 16]. These critical magnetic field values ($H_c$) correspond to the maxima in $dM/dH$. It is important to note that a small but clear hysteresis is observed at $H_{c2}$ and $H_{c3}$ which was not noticed in previous studies [12, 16]. This establishes first-order nature of these transitions. With increasing temperature, the width of the hysteresis decreases, as inferred by a comparison of the curves for 2 and 5 K. Another new observation to be noted is that, below $H_{c2}$, there is a small convex-shaped curvature which indicates the existence of one more magnetic anomaly (Fig. 4a). We have plotted $dM/dH$ in Fig. 4d, which clearly reveals an additional weak feature around 6 kOe ($H_{c1}$). The transition at $H_{c1}$ does not show any hysteresis. At 12.5K (Fig. 3c, right y axis), a clear magnetic-field-induced transition is seen in $M$ $vs$ $H$ above 50 kOe, but a careful look at the derivative curve reveals the existence of more than one field-induced transition; a small but clear anomaly can be seen at a very low field (inset of Fig. 4c). The magnetization value at the highest measured magnetic-field at low temperatures is found to be $\sim 9\mu_B/f.u.$, which agrees well with the neutron diffraction results i.e. the sum of $Dy^{3+}$ and $Ni^{2+}$ ions magnetic moments at low temperature [15]. It is worth noting that magnetic-field-induced multiple transitions for $1D$ chain systems $Ca_3Co_2O_6$ are of great interest [see, for instance, references 29 and 30].

In order to explore $ME$ coupling, we have performed isothermal $\varepsilon'$ $vs$ $H$ measurements at different temperatures up to 140 kOe and presented in the form of $\Delta\varepsilon'$, where $\Delta\varepsilon'= [(\varepsilon'_H-\varepsilon'_{H=0})/\varepsilon'_{H=0}]$ in Fig. 4e, f and g at the same temperatures as for $M$ $vs$ $H$ curves, i.e., at 2, 5 and 12.5K. Consistent with the behavior observed in $M$ $vs$ $H$ data, two transitions with hysteresis are observed at 2 and 5K near $H$ =40 and 60 kOe; it is notable that a small anomaly is seen near $H_{C1}$ (marked with vertical dotted line) which is clearly visible in the plot of d$\Delta\varepsilon'$/d$H$ for 2 K (Fig. 4h). Even at 12.5K, the low-field transition is more prominent in $\Delta\varepsilon'$ $vs$ $H$ plot (Fig. 4g). This shows that $\varepsilon'$ $vs$ $H$ measurements are very sensitive for detecting magnetic-field-induced changes [30, 31], if there is a $ME$ coupling. Similar behavior has been also observed in the well-known multiferroic compound $CuCrO_2$ [31] in which magnetic-field-induced transition was observed only in $\varepsilon'$ $vs$ $H$, even for single crystals. In fact, if one looks at d$M$/d$H$ data carefully for $Er_2BaNiO_5$ [16], three magnetic-field-induced transitions can actually be inferred supporting our observation. Another interesting observation is that the value of $\Delta\varepsilon'$ is quite high and is comparable to that known for $TbMnO_3$ multiferroics (along $a$-axis) [1] and larger than that known for other $ME$ materials [30, 32]. In fact it is also much higher than that observed for manganites [33]. Interestingly, a maximum value of $\Delta\varepsilon'$ (~4%) is observed for intermediate temperatures, e.g., at 12.5 and 15K, and then the magnitude decreases at higher temperatures. But no $\Delta\varepsilon'$ is observed above $T_N$ i.e. at 80K, in the paramagnetic state.

Remnant polarization as a function of $H$ was also measured, for instance, at 7K. The sample was cooled with E= 400kV/m from paramagnetic/paraelectric region to 7K as described for $P$ $vs$ $T$ measurements and then $P$ $vs$ $H$ was measured (100Oe/sec). For a direct comparison of different order parameters (spin/charge) as a function of $H$, we have shown $H$-dependent $M$, $\varepsilon'$ and $P$ in Fig. 5 a-c at 7K. Fig. 5c shows that $P$ decreases with increasing $H$, exhibiting a change of curvature in the vicinity of magnetic-fields where there are distinct changes in other $H$-dependent measurements. A gigantic change in $P$ (1500%) (see Fig. 5c) is observed with $H$ and, at $H_{c2}$, even a change of sign occurs. The large change in $P$ at the magnetic-field induced transition is also observed for other well known multiferroics [1, 31]. Recently, a similar behavior i.e. a large change in $P$ and a change of $P$ sign at the critical field was reported in GdMn$_2$O$_5$ [34]. All these results show one-to-one correlation with magnetic properties which again prove the existence of $ME$ coupling in this compound.

## IV.    DISCUSSIONS

As mentioned above, three magnetic anomalies are observed below $T < T_{max}$, both in $M$ $vs$ $H$ and $\varepsilon'$ $vs$ $H$. The presence of these magnetically induced transitions could be related to different relative orientations of Dy$^{3+}$ and Ni$^{2+}$ magnetic moments, based on the knowledge existing for varying temperature as elaborated below.

Ni$^{2+}$ moments are coupled to each other antiferromagnetically along $a$-axis. The neutron diffraction results [15] show that, for Dy$_2$BaNiO$_5$ at low $T$ ($T$=1.5K), the Ni$^{2+}$ moments align towards $c$-axis ($ac$ plane) i.e. towards the direction of Dy$^{3+}$ moments by making a small angle ($\theta$= 17°) with $c$-axis. This angle of Ni moment with $c$-axis is very small relative to that for the rest of the members of this series [15]. The rotation of Ni$^{2+}$ magnetic moments with decreasing temperature is strongly $R$-dependent and this is due to the different anisotropy of $R$ ions [15, 26]. In the case of Dy$_2$BaNiO$_5$, at low temperatures, Ni$^{2+}$ moments rotate towards $c$-axis due to large magnetic moment of Dy$^{3+}$ and strong $J_{Ni-Dy}$ exchange interactions [15]. Temperature evolution of Ni$^{2+}$ moments [15] revealed that $\theta$ is different in different regions. For instance, close to $T_N$, it is negative with respect to Dy$^{3+}$ moment and fluctuating; however, with a further decrease in $T$, there is a crossover of $\theta$ from a negative to a positive value and this crossover temperature is close to $T_{max}$. At very low $T$ (close to 10K), there is again a small curvature in $\theta$ [see Fig. 6b of Ref. 15]. Moreover, this crossover of sign can also be seen clearly in d$\chi$/d$T$ *vs* $T$ plot (Fig. 1a) near 38 and 10K.

Now viewing all the results - *M vs T*, $\varepsilon'$ *vs T* and *P vs T* behavior- together with *T* evolution of Dy$^{3+}$ magnetic moments and Ni$^{2+}$ angle with *c*-axis, we can infer approximately four different temperatures regions: (I) $T$<12.5K; (II) between 12.5K and $T_{max}$; (III) from $T_{max}$ to 50K; and (IV) from 50K to 60K. No structural change had been detected by neutron diffraction studies down to low temperatures [15]; only *a* axis shows discontinuity at 3*D AFM* ordering temperature for Tb, Ho and Tm, indicating the presence of magnetostriction effect also [8]. However, no such discontinuity is known in this compound, and so the magnetism is solely responsible for the appearance of polarization. The Ni$^{2+}$ and Dy$^{3+}$ moments are canted and the magnetic structure is commensurate having single wave vector *k(1/2, 0, 1/2)* with point-group of little group 2/*m* [15]. On the basis of magnetic symmetry analysis, Nénert *et al.* [24] determined that the magnetic point group is 2/*m'*, which is favorable for electric polarization and linear *ME* coupling [35].

We can in principle extend the arguments proposed to explain the temperature-induced anomalies in terms of the rotation of Ni moments, to the magnetic-field induced transitions as well. As stated earlier, *M vs H* data at low *T* (2 and 5K, region I) show a notable feature at small *H* (<$H_{c2}$) without any hysteresis. At $H_{c2}$ and $H_{c3}$, *M* increases rather sharply giving rise to hysteresis. We speculate that this sharp increase at $H_{c2}$ and $H_{c3}$ could be related to Ni$^{2+}$ moment aligning nearly towards $Dy^{3+}$ moment direction (c-axis) i.e. $Ni^{2+}$ and $Dy^{3+}$ aligned ferromagnetically. Careful neutron diffraction measurements as a function of magnetic-field are urgently warranted to get more precise information about the relative orientation of Ni moment with respect to Dy moment and to correlate with the features reported in this article. At $T{\geq}T_{max}$ (region III), we did not observe any magnetic-field-induced transitions, neither in $\Delta\varepsilon'$ *vs H* nor in *M vs H*.

This sample is highly insulating in the magnetic ordering temperature region. Therefore, the observed magnetoelectric coupling is the intrinsic property of this sample and does not arise from any extraneous factors like magnetoresistance or leakage current. We would like to mention that, in order to rule out grain boundary contributions or contact effects, we have carried out impedance measurements at 50 K and obtained Nyquist plot, that is, a plot of real part versus imaginary part. We found that there is no complete semicircular arc which is expected for semiconductors/relaxor materials [36]; also, we found that the plot is not sensitive to any variation of applied ac bias voltage (1 mV to 2V), which will not be the case if the observed features arises due to external contributions [37].

## V. CONCLUSIONS

The *ME* coupling in Haldane spin chain 1*D* nickelate $Dy_2BaNiO_5$ is explored by *T*- and *H*-dependence of dielectric constant and electric polarization. Similar to those in *M vs T*, anomalies are observed in $\varepsilon'$ (*tan δ*) and *P vs T* data in the magnetically ordered state, which suggests multiferroic behavior and strong *ME* coupling in this compound. An additional outcome of this work is that there are actually three (rather than two) magnetic-field-induced transitions (say, at 2 K near ~6, 40 and 60 kOe), with the data establishing that dielectric data is sensitive to detect magnetic transitions, if there is a magnetoelectric coupling. We interpret that the temperature and magnetic-field-induced transitions depend on relative orientations of $Ni^{2+}$ and $Dy^{3+}$ magnetic moments. Finally, the topic of successive magnetic transitions resulting in multiferroicity in $S=1$ systems is of great current interest [38] and, in that sense, the present results gain importance, providing another material of this kind for further work on this topic.

**ACKNOWLEDGMENTS**

The author K. S. is thankful to Dr. C. S. Yadav for his help during initial stages and also to all technical staff for their contributions in the development of this set up.

**Figure Captions:**

Fig. 1. Temperature dependence of **(a)** zero-field-cooled (*zfc*) magnetic susceptibility ($\chi$) measured in 5 kOe (left y-axis) for $Dy_2BaNiO_5$. $d\chi/dT$ is plotted on right y-axis axis; **(b)** Temperature dependence of inverse susceptibility; the line represents Curie-Weiss fitting in the paramagnetic region.

Fig. 2. Temperature (*T*) dependence of **(a)** heat capacity (*C*) divided by temperature, and **(b)** d(*C/T*)/d*T*, for $Dy_2BaNiO_5$.

Fig. 3. Temperature dependence of **(a)** dielectric constant measured at 30kHz, **(b)** *tan$\delta$* (left y-axis) and d(*tan$\delta$*)/d*T* (right y-axis), and **(c)** remnant polarization measured during warming for $Dy_2BaNiO_5$. The line through the data points in (c) represents the polynomial fit (see text for more details) as a guide to the eyes. Insets in (a) show $d^2\varepsilon'/dT^2$ (left, lower side) and dielectric constant on expanded scale (right, upper side) and inset in (b) shows *tan$\delta$* on expanded scales at low temperatures.

Fig. 4. Magnetic-field dependence of magnetization (**a, b** and **c**), and dielectric constant (**e, f,** and **g**) for $Dy_2BaNiO_5$ at different temperatures. **(c)** right y axis and **(d)** show d*M*/d*T* curves at 12.5 and 2K respectively. **(h)** d$\Delta\varepsilon'$/d*H* at 2K. Inset in (c) and (g) show d*M*/d*T* and dielectric constant at 12.5K at low magnetic fields.

Fig. 5. Magnetic-field dependence of magnetization **(a)**, dielectric constant **(b)** and electric polarization **(c)** of $Dy_2BaNiO_5$ at 7K.

**Fig. 1.**

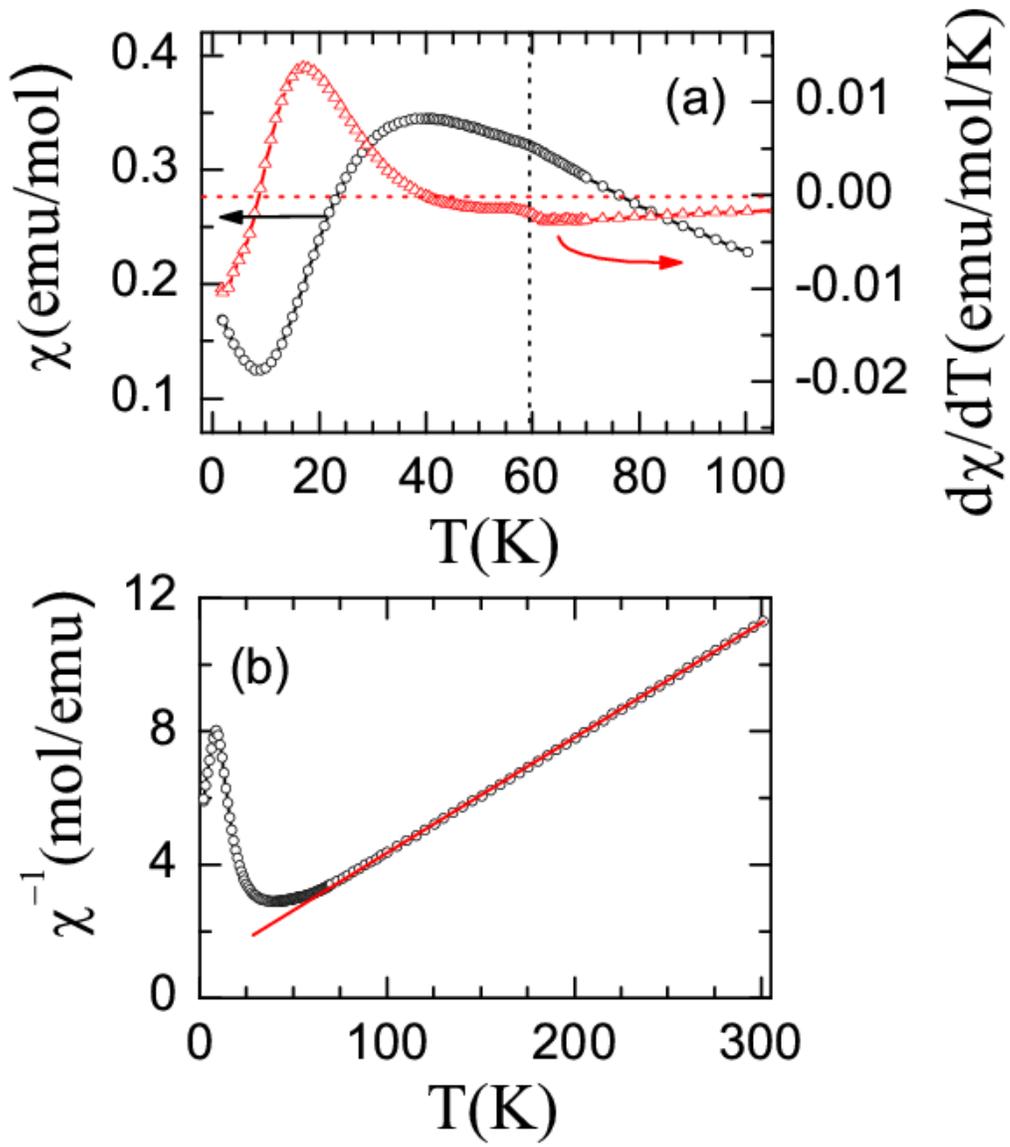

**Fig. 2.**

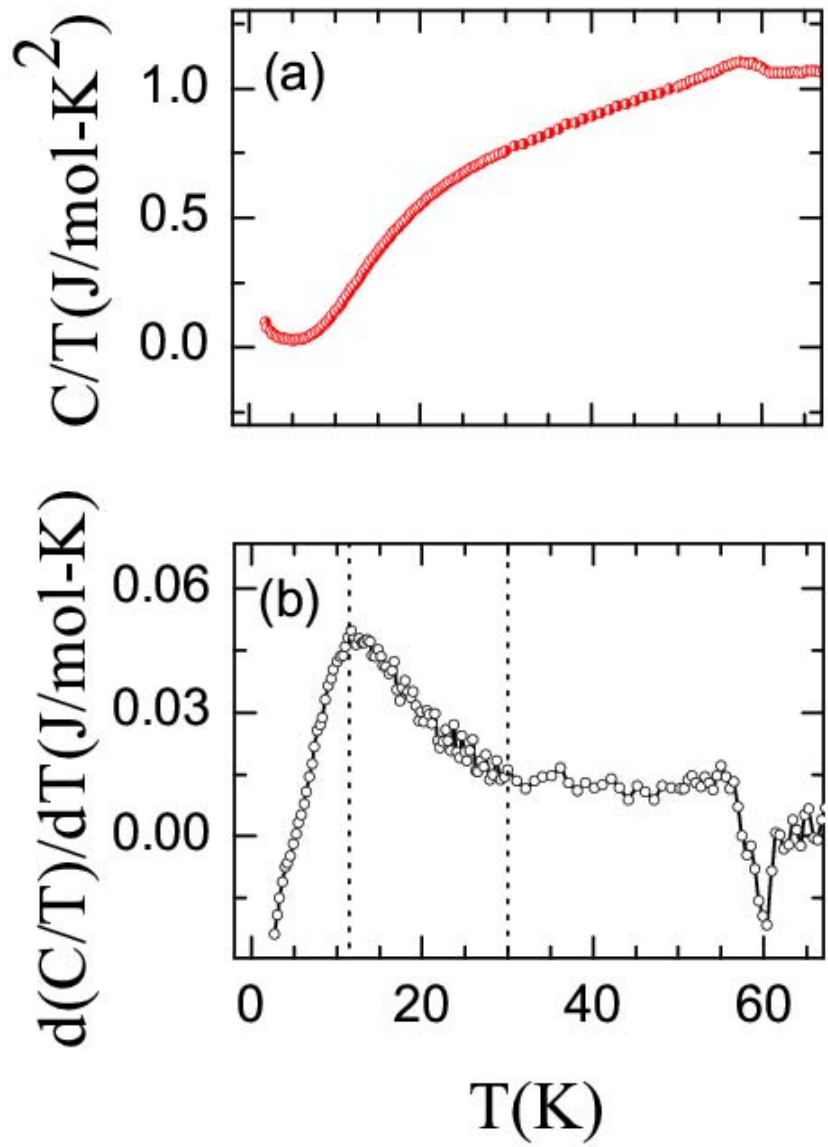

**Fig. 3.**

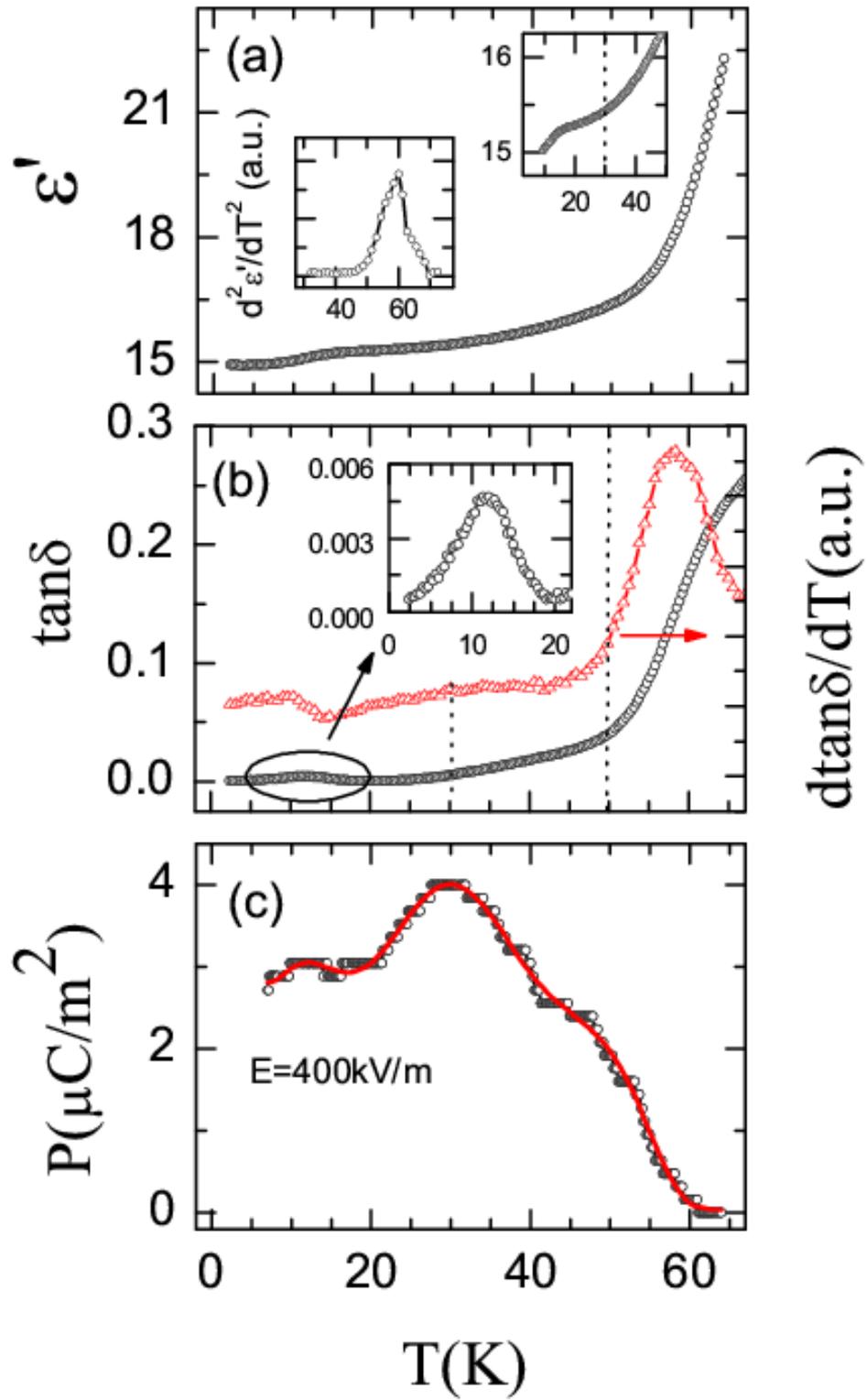

**Fig. 4.**

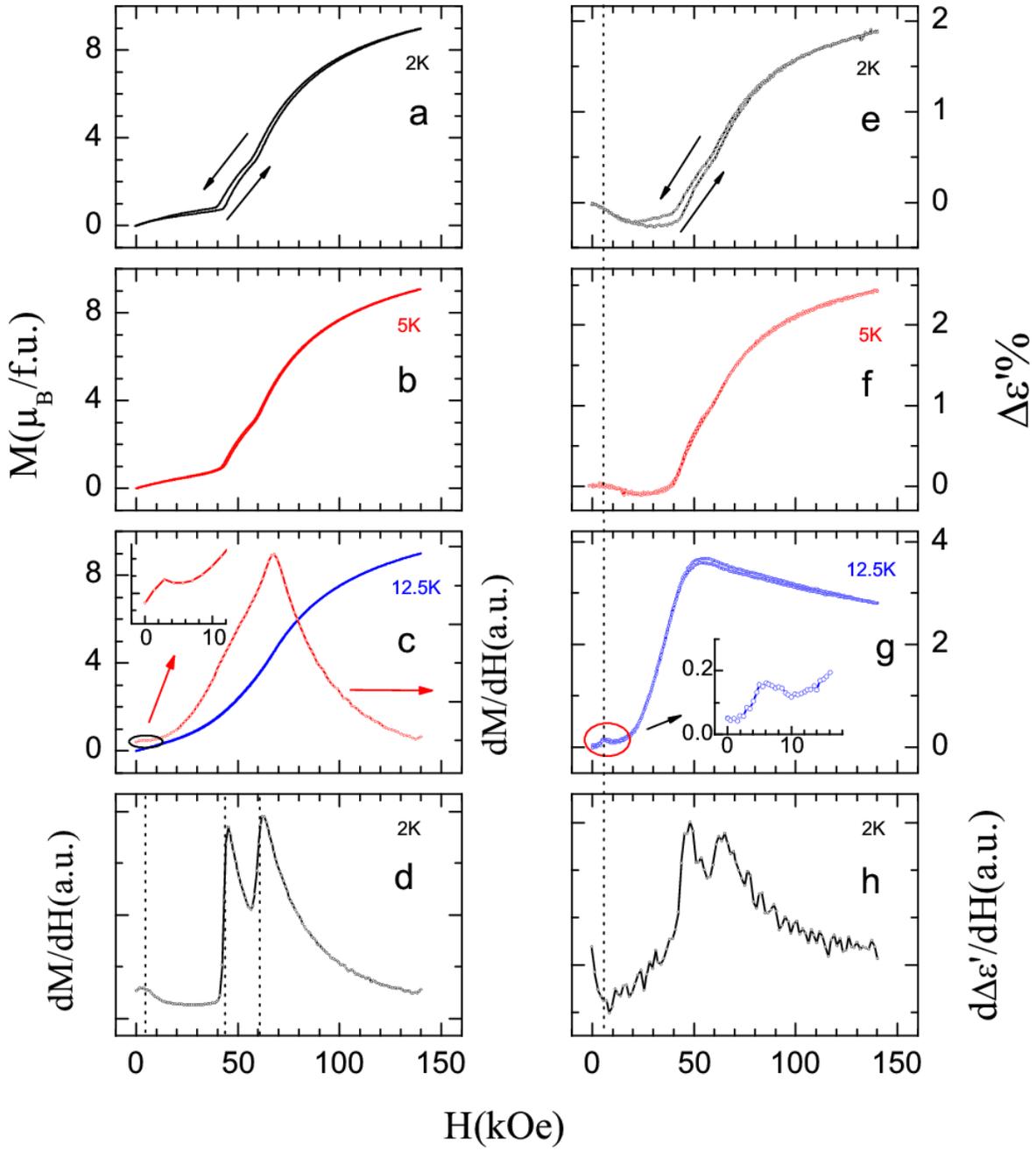

**Fig. 5.**

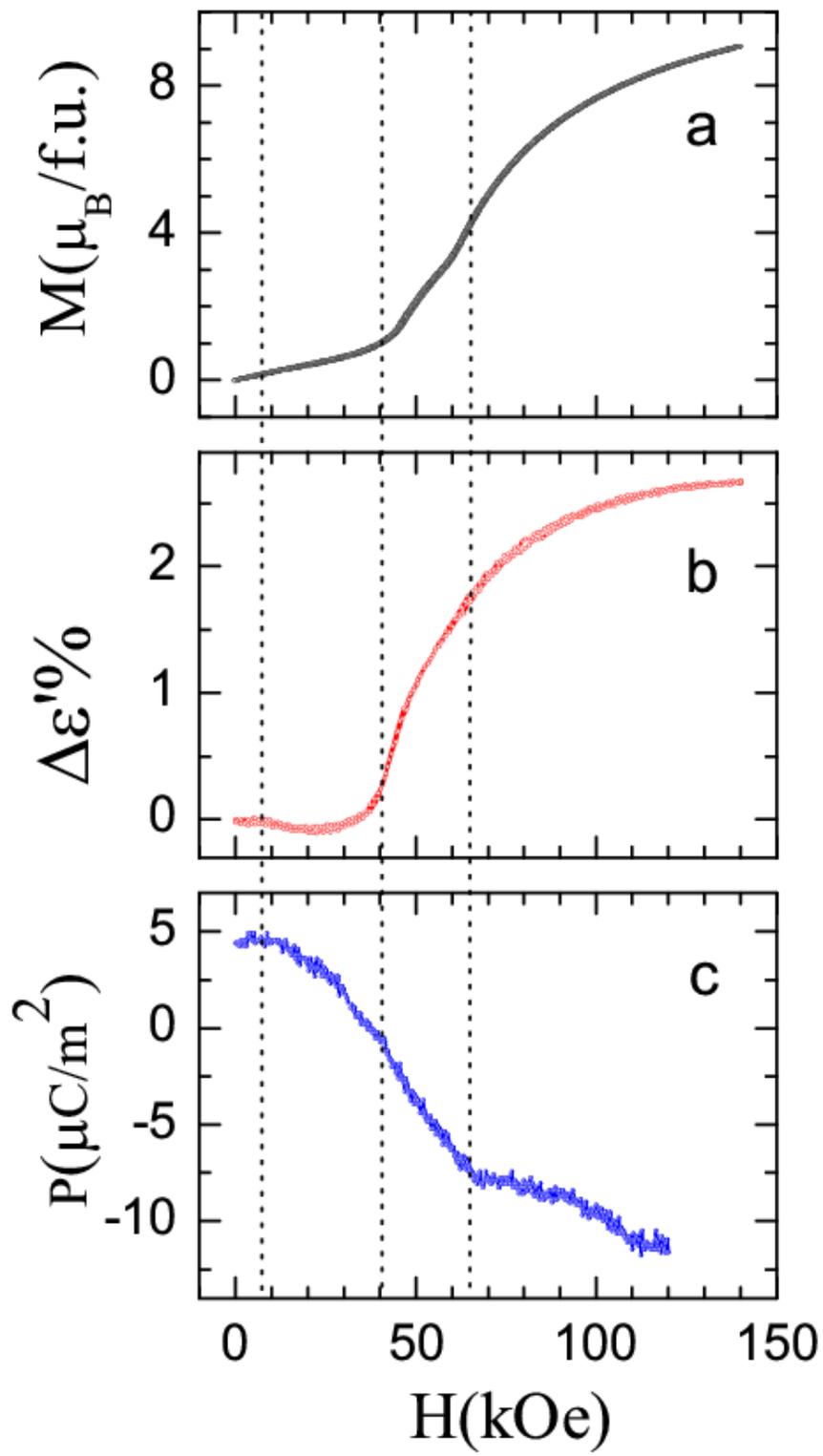